\documentclass[arps,twocolumn,floats,prd,nofootinbib,showpacs,showkeys]{revtex4}

\usepackage{dsfont}
\usepackage{amsmath}
\usepackage{amssymb}
\usepackage{graphicx}
\usepackage{verbatim}
\usepackage{natbib}
\usepackage{eucal}
\usepackage{calligra}

\DeclareMathAlphabet{\mathscr}{OT1}{pzc}%
                                 {m}{it}

\newcommand{\be}{\begin{equation}}
\newcommand{\ee}{\end{equation}}
\newcommand{\bes}{\begin{equation*}}
\newcommand{\ees}{\end{equation*}}
\newcommand{\bea}{\begin{eqnarray}}
\newcommand{\eea}{\end{eqnarray}}
\newcommand{\beas}{\begin{eqnarray*}}
\newcommand{\eeas}{\end{eqnarray*}}

\newcommand{\g}{\mathscr{g}}
\newcommand{\gbar}{\overline{\g}}
\newcommand{\gvir}{\gbar_{\rm vir}}

\newcommand{\msun}{M_{\odot}}
\newcommand{\ten}{10^{10}}
\newcommand{\eleven}{10^{11}}
\newcommand{\twelve}{10^{12}}
\newcommand{\thirteen}{10^{13}}
\newcommand{\fourteen}{2 \times 10^{14}}

\begin{document}

\title{Halo Scale Predictions of Symmetron Modified Gravity}
\author{Joseph Clampitt, Bhuvnesh Jain, Justin Khoury}
\email{clampitt@sas.upenn.edu}
\affiliation{Dept. of Physics and Astronomy, University of Pennsylvania}
\keywords{Cosmology and Extragalactic Astrophysics, General Relativity and Quantum Cosmology}

\begin{abstract}
We offer predictions of symmetron modified gravity in the neighborhood of realistic dark matter halos. The predictions for the fifth force are obtained by solving the nonlinear symmetron equation of motion in the spherical NFW approximation. In addition, we compare the three major known screening mechanisms: Vainshtein, Chameleon, and Symmetron around such dark matter sources, emphasizing the significant differences between them and highlighting observational tests which exploit these differences. Finally, we demonstrate the host halo environmental screening effect (``blanket screening'') on smaller satellite halos by solving for the modified forces around a density profile which is the sum of satellite and approximate host components.
\end{abstract}

\date{\today}

\maketitle

\section{Introduction}

The observed acceleration in the expansion of the universe can arise from a dark energy component or from a departure of gravity from general relativity (GR) on cosmological scales. One way to distinguish between the two possibilities is to consider the growth of perturbations. For modified gravity (MG) theories, the relation of the expansion history to the growth of perturbations is specific to every model. In the quasi-static, Newtonian linear regime, several authors have parameterized the growth of perturbations with $g(k,z) \equiv G/G_N$ and $\eta(k,z) \equiv \psi/\phi$ (e.g. \citep{jaietal08, sch08, zhaetal09}).

Laboratory and solar system constraints (see \citep{wil05} and referrences therein) require any viable MG theory to have some mechanism by which it mimics the predictions of GR within Milky Way-size halos. Such ``screening'' mechanisms (see \citep{jaietal10} for a review) generally determine the deviations from GR based on the local density: in high density environments the scalar force is suppressed, while in low density environments it can be of approximately gravitational strength. Two such mechanisms have been extensively studied in the literature: the chameleon screening \citep{khoetal04} of $f(R)$ theories \citep{huetal07} and Vainshtein screening \citep{vai72} of higher dimensional (e.g., Dvali-Gabadadze-Porrati (DGP) gravity \citep{dvaetal00}) and Galileon models \citep{nicetal09}. Recent work has included detailed simulations which are necessary because of the nonlinearity inherent in how GR is recovered inside the Milky Way \citep{oya08, sch09, chaetal09}.

In this work we explore the symmetron model of \citep{hinetal10} (see also \citep{olietal08} and \citep{pie05}), which exploits a novel screening mechanism similar in part to chameleon screening but with key differences. The symmetron has a vacuum expectation value (VEV) that is large in low density environments and small in high density environments. Symmetron screening then relies on a coupling to matter that is proportional to the VEV, thus suppressing the scalar force in high density environments. Recent work has focused on symmetron cosmology, including the evolution of the symmetron field through various cosmological epochs \citep{hinetal11}, as well as its effect on linear \citep{braetal11} and nonlinear \citep{davetal11b} structure formation.

Tests of gravity on linear scales have some limitations. The $g, \eta$ parameterization is only valid on scales smaller than the superhorizon regime and larger than the nonlinear regime. At high redshift there is a different problem: since MG models recover GR at high redshift for consistency with CMB and Nucleosynthesis observations, effects of enhanced forces are manifested only at late times. Thus, even if observations are made late enough to be within the MG era, the signal has had limited time to accumulate.

In this study we consider tests on scales within and outside virial radii of dark matter halos modeled with the Navarro-Frenk-White (NFW) \citep{navetal97} profile. In this regime the predicted deviations due modified gravity can be significantly larger than in the linear regime (measurement errors and systematic uncertainties need to be taken into account but will not be considered here). As highlighted by \citep{huietal09} and \citep{jaietal11}, astrophysical tests in this regime can provide effective tests of chameleon theories. We will calculate the predicted deviations for symmetron theories.

In \S~II we describe the symmetron theory and our method for calculating modified forces around NFW halos. \S~III contains our isolated halo results as well as a comparison of the various screening mechanisms. In \S~IV we model and show results for forces on test particles in two-body host-satellite systems. We conclude in \S V.

\section{Force Profiles of NFW Halos}
\label{sec:theory}

\subsection{Symmetron Theory}

In the Einstein frame we can describe the gravitational forces as GR with an additional ``fifth force,'' mediated by the symmetron field, $\phi$. For GR we have the usual Poisson equation: $\nabla^2 \Psi_{\rm N} = 4\pi G\rho$, leading to
\be \label{eq:nforce}
|F_{\rm N}| = \frac{{\rm d}\Psi_{\rm N}}{{\rm d}r} = \frac{G M(<r)}{r^2} \, .
\ee

The symmetron equation of motion in the presence of non-relativistic matter \citep{hinetal10} is
\be \label{eq:gen-eom}
\Box \phi = \frac{\partial V}{\partial \phi} + \rho \frac{\partial A}{\partial \phi} \equiv \frac{\partial}{\partial \phi} V_{\rm eff} \, ,
\ee
where
\be \label{eq:potential}
V (\phi) = -\frac{1}{2} \mu^2 \phi^2 + \frac{1}{4} \lambda \phi^4 \, ,
\ee
and
\be \label{eq:coupling}
A(\phi) = 1 + \frac{\phi^2}{2M_{\rm s}^2} + {\cal O}\left(\frac{\phi^4}{M_{\rm s}^4}\right)\,.
\ee
Note that the relevant field range is $\phi \ll M_{\rm s}$, such that any ${\cal O}(\phi^4/M_{\rm s}^4)$ terms in $A(\phi)$ can be consistently neglected. The potential $V(\phi)$ comprises the most general renormalizable form invariant under the $\mathds{Z}_2$ symmetry $\phi\rightarrow -\phi$.
The coupling to matter $\sim \phi^2/M_{\rm s}^2$ is the leading such coupling compatible with the symmetry. The model involves two mass scales, $\mu$ and $M_{\rm s}$, and one positive dimensionless coupling $\lambda$. The mass term is tachyonic, so that the $\mathbb{Z}_2$ symmetry $\phi\rightarrow -\phi$ is spontaneously broken. The effective potential of Eq.~(\ref{eq:gen-eom}) is
\be
V_{\rm eff}(\phi)={1\over 2}\left({\rho\over M_{\rm s}^2}-\mu^2\right)\phi^2+{1\over 4}\lambda\phi^4\,.
\ee
Whether the quadratic term is negative or not, and hence whether the $\mathbb{Z}_2$ symmetry is spontaneously broken or not, depends on the local matter density. 

The screening mechanism works roughly as follows: in vacuum or in large voids, where $\rho\simeq 0$, the potential breaks reflection symmetry spontaneously, and the scalar acquires a VEV $|\phi| = \phi_0\equiv \mu/\sqrt\lambda$; in regions of high density, such that  $\rho > M_{\rm s}^2\mu^2$, the effective potential no longer breaks the symmetry, and the VEV goes to zero. Meanwhile, to lowest order the symmetron-matter coupling is $\sim\rho \phi^2/M_{\rm s}^2$. Fluctuations $\delta\phi$ around the local background value $\phi_{\rm VEV}$, which would be detected by local experiments, couple to density as
\be 
\sim{\phi_{\rm VEV}\over M_{\rm s}^2}\delta\phi \ \rho\,.
\label{coupling}
\ee
In particular, the coupling is proportional to the local VEV.  In high-density environments where the symmetry is restored, the VEV should be near zero and fluctuations of $\phi$ do not couple to matter. In less dense environments, where $\rho < M_{\rm s}^2\mu^2$ and the symmetry is broken, the coupling turns on.

For a static-spherically symmetric source, Eq.~(\ref{eq:gen-eom}) becomes 
\be \label{eq:eom}
\frac{{\rm d}^2 \phi}{{\rm d}r^2} = -\frac{2}{r}\frac{{\rm d}\phi}{{\rm d}r} + \left(\frac{\rho}{M_{\rm s}^2}-\mu^2 \right)\phi + \lambda \phi^3 \, .
\ee
We set the parameters as in \citep{hinetal10}: $M_{\rm s} = 10^{-3} M_{\rm Pl}$ satisfies solar system constraints while still allowing for order unity deviations elsewhere. Also, $\mu = \sqrt{\rho_c} / M_{\rm s}$ and $\lambda = (\mu / \phi_0)^2$, where $\rho_c$ is the average cosmological density today and $\phi_0$ is the background value of the field. Note that these parameter choices correspond to $\mu \sim {\rm Mpc}^{-1}$, constraining symmetron effects to $\sim$ Mpc distances.

As in \citep{hinetal11}, the symmetron-mediated force $F_\phi$ relative to the Newtonian force $F_{\rm N}$ between two test masses in vacuum is set by the symmetry-breaking value $\phi_0$:
\be \label{eq:ratio}
\frac{F_\phi}{F_{\rm N}} = 2M_{\rm Pl}^2 \left(\frac{{\rm d}\ln A}{{\rm d}\phi}\bigg\vert_{\phi_0} \right)^2 \simeq 2\left(\frac{\phi_0 M_{\rm Pl}}{M_{\rm s}^2}\right)^2\, .
\ee
If the scalar-mediated force is to be comparable to gravity in vacuum, then we
must impose $\phi_0/M_{\rm s}^2 \sim 1/M_{\rm Pl}$, that is,
\be
\phi_0\equiv \frac{\mu}{\sqrt{\lambda}} =  g \frac{M_{\rm s}^2}{M_{\rm Pl}}\, ,
\label{vev}
\ee
where $g \sim {\cal O}(1)$. To be precise, it follows from Eq.~(\ref{eq:ratio}) that $g$ measures the strength of the scalar force in vacuum relative to gravity:
$F_\phi = 2g^2F_{\rm N}$. For comparison to $f(R)$ and DGP theories, for which the fifth force is at most 1/3 $F_{\rm N}$, we will set $g = 1/\sqrt{6}$; otherwise we choose $g = 1$, which is still consistent with solar system tests, as shown in \citep{hinetal10}. Note that Eq.~(\ref{eq:ratio}) has no dependence on the test bodies involved. Extended mass distributions affect the scalar and Newtonian forces differently so that solving for the Newtonian potential and scalar field profile is required in order to evaluate the ratio. The ratio of forces on a test mass in the neighborhood of such an extended distribution is
\be \label{eq:ratio2}
\frac{F_{\phi}}{F_{\rm N}} = \frac{(\phi/M_{\rm s})(\nabla \phi /M_{\rm s})}{\nabla \Psi_{\rm N}} \, .
\ee

\subsection{NFW Halos}

\begin{figure}
\includegraphics[width=0.85\columnwidth]{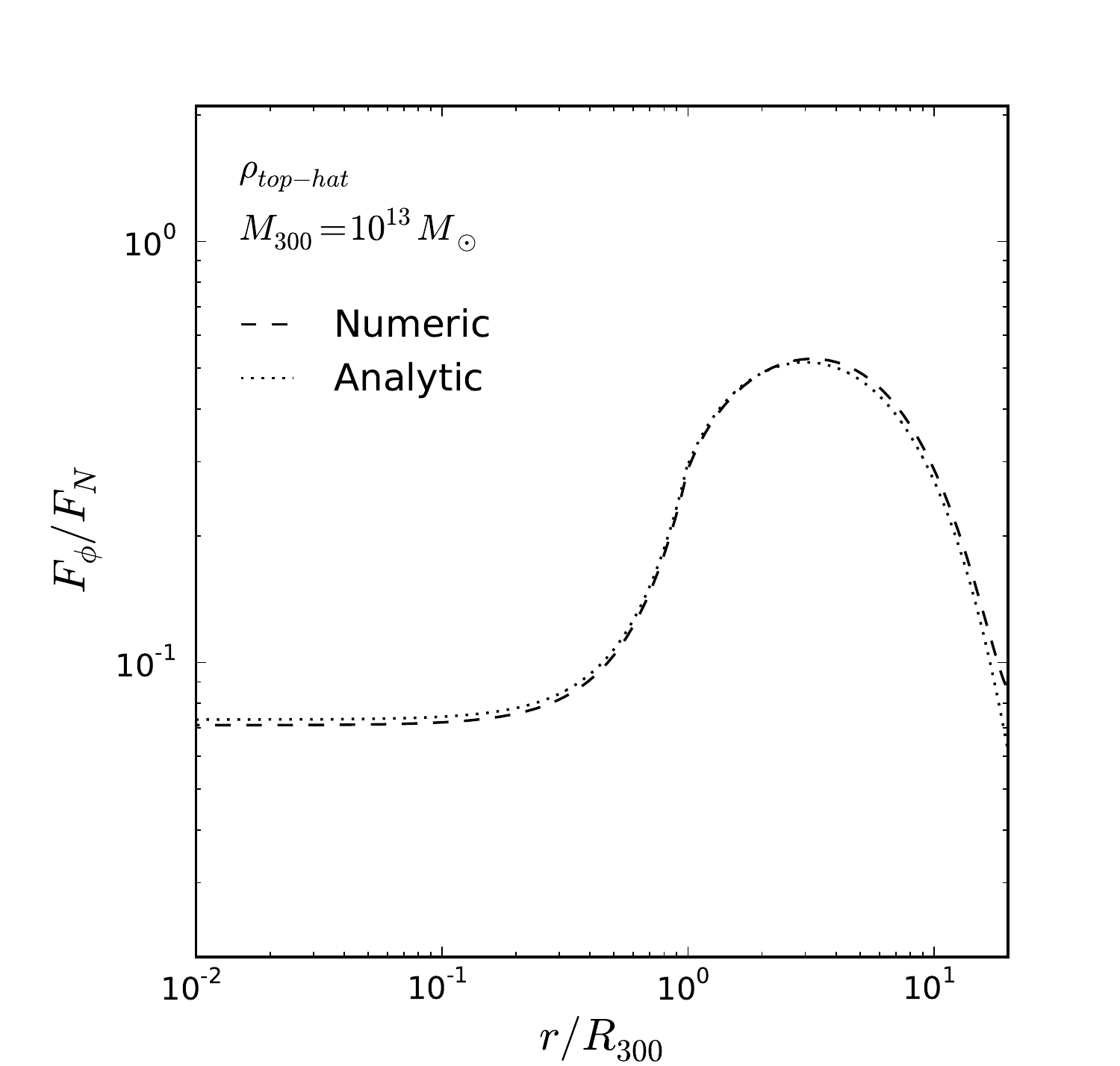}
\caption{Force deviation $F_{\phi}/F_{\rm N}$ for a top-hat density profile of total mass $\thirteen \msun$ calculated numerically (dashed line) and analytically (dotted line). We find similar agreement between the two methods throughout the mass range $10^{10}-\fourteen \msun$.}
\label{fig:tophat}
\end{figure}

We consider gravitational forces in the neighborhood of NFW halos \citep{navetal97}, whose density profiles are a good fit to those of stacked simulated halos. The (untruncated) density is
\be \label{eq:rho}
\rho_{\rm NFW} (r) = \frac{4\rho_{\rm s}}{\frac{r}{r_{\rm s}} \left(1+\frac{r}{r_{\rm s}}\right)^2} \;,
\ee
where $\rho_{\rm s}$ and $r_{\rm s}$ are parameters that depend on halo mass (see below). We define the mass $M_{300}$ of each halo as that enclosed within the virial radius, $R_{300}$. This is the radius at which the average density enclosed is 300 times the critical density $\rho_c$. The concentration connects the virial radius to the scale radius, $c = R_{300} / r_{\rm s}$. We take it to be a function of halo mass, 
\be \label{eq:c}
c = 9 \left(\frac{M_{300}}{3.2 \times 10^{12}M_{\odot}/h} \right)^{-0.13} \, ,
\ee
as found in \citep{buletal01}. The mass and concentration definitions are chosen to allow comparison of our results with the chameleon model of \citep{sch10}. We note that the profile itself will change somewhat with the modified forces.

The mass enclosed within radius $r$ is
\be \label{eq:mass}
M (<r) = M_{300} \frac{F(c \; r/R_{300})}{F(c)} \, ,
\ee
where $F(x) = \ln{(1+x)} - x/(1+x)$. We use Eq.~(\ref{eq:mass}) in (\ref{eq:nforce}) to solve for the Newtonian force. In order to obtain the symmetron profile, it is useful to define a dimensionless scalar field, $\psi \equiv \phi / \phi_0$ whose equation of motion follows from Eq.~(\ref{eq:eom}):
\be \label{eq:new-eom}
\frac{{\rm d}^2 \psi}{{\rm d}r^2} = -\frac{2}{r}\frac{{\rm d}\psi}{{\rm d}r} + \left(\frac{\rho}{M_{\rm s}^2}-\mu^2 \right)\psi + \mu^2 \psi^3 \, .
\ee
We obtain the radial profile by substituting Eq.~(\ref{eq:rho}) in ~(\ref{eq:new-eom}) with boundary conditions
\be \label{eq:new-bc}
\left.\frac{d\psi}{dr}\right|_{r=0} = 0, \; \; \; \psi (r \rightarrow \infty) = 1 \, .
\ee
We use a shooting algorithm to solve this nonlinear equation, tuning the boundary condition $\psi(r=0)$ such that the field stays within 1\% of $\psi=1$ at large $r$ for at least 25 virial radii. Note that Eqs.~(\ref{eq:new-eom}), (\ref{eq:new-bc}) depend only on the theory parameters $\mu$ and $M_{\rm s}$; they are independent of $\phi_0$ and $\lambda$. The only effect of changing $\phi_0$ is to set the overall amplitude of the symmetron profile and therefore the amplitude of the ratio of forces.

For comparison to the work of \cite{hinetal10} and as a check of our numerical solutions, we also solve for the force deviation in the case of a top-hat density profile
\be \label{eq:tophat}
\rho_{\rm top-hat}(r) = \begin{cases} \rho_0, & r < R_{\rm vir} \\ 0, & r > R_{\rm vir} \, , \end{cases}
\ee
where $\rho_0 = 300 \rho_{\rm c}$. For the top-hat profile, specifying the mass and density fixes $R_{300}$. To obtain an analytic solution, we approximate the symmetron equation of motion as quadratic around the appropriate minimum inside and outside the object (see \cite{hinetal10} for details) resulting in
\bea \label{eq:tophat-approx}
\phi_{\rm in} (r) & = & A\frac{R}{r}\sinh\left({r\sqrt{\frac{\rho}{M_{\rm s}} - \mu^2}}\right) \nonumber \\
\phi_{\rm out}(r) & = & B\frac{R}{r}e^{-\sqrt{2}\mu r} + \phi_0 \, ,
\eea
and solve for $A$ and $B$ by matching at the boundary. The exact solution is obtained using our shooting algorithm as in the NFW case. Figure \ref{fig:tophat} shows the equivalence of these methods for a top-hat mass $\thirteen \msun$. We find similar agreement between the two methods throughout the mass range $10^{10}-\fourteen \msun$.

We can define $\gvir$ as an average of the force deviation over the virial radius of a halo (see \citep{sch10} for details). This quantity can be determined from both theory and observations. From the theory we have calculated
\be \label{eq:gvir}
\gvir = \frac{\int r^3 \rho(r)\:F_{\rm N} (1+F_{\phi}/F_{\rm N})\:dr}{\int r^3 \rho(r)\:F_{\rm N}\:dr} \, ,
\ee
where the integral is over the virial radius $R_{300}$. This is straightforward to compare to observations, which yield
\be \label{eq:gvir-obs}
\gvir = (M_{300, {\rm dyn}}/M_{300})^{5/3} \, ,
\ee
where $M_{300, {\rm dyn}}$ is a dynamical mass, and $M_{300}$ is the ``true'' or lensing mass.

As will be seen later, the deviation from GR is significant out to $\sim 10$ times the virial radius. Therefore we define a second average $\bar{\g}_d$ exactly as in Eq.~(\ref{eq:gvir}), except the integrals are taken over distances $d = 0.5, \; 1, \; 4,$ and $10 \; R_{300}$.

\section{Results for Isolated Halos}

\begin{figure*}
\centering
\includegraphics[width=6.5in]{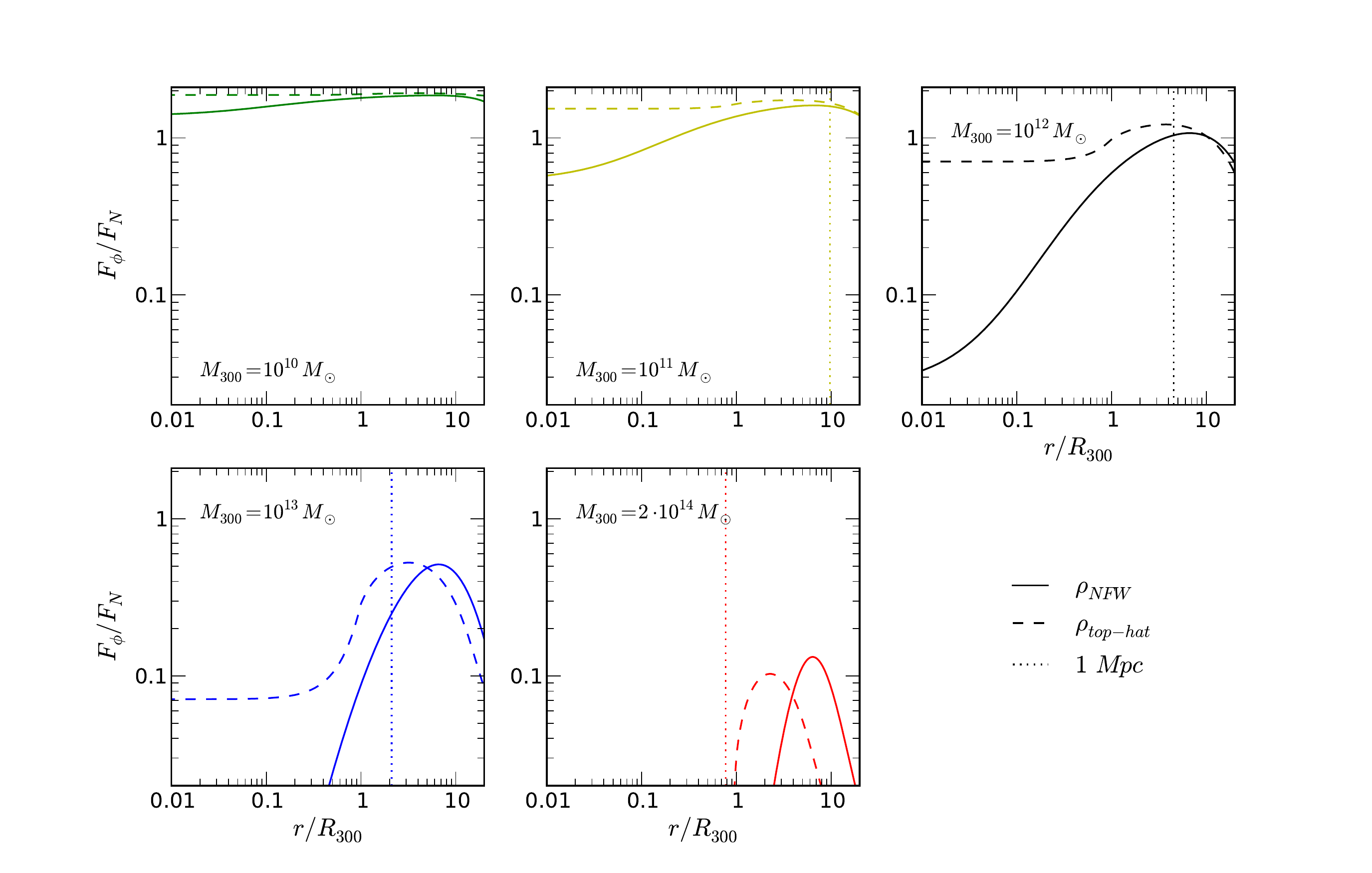}
\caption{Ratio $F_{\phi}/F_{\rm N}$ of the symmetron mediated force to the GR force for halos of different masses with field symmetry breaking value $\phi_0 = M_{\rm s}^2/M_{\rm Pl}$ (i.e. $g=1$). The halos are modeled with NFW (solid line) and spherical tophat (dashed line) profiles. Note that the ratio is plotted vs. radius in units of each halo's virial radius, $R_{300}$. The background compton wavelength, $\lambda_{\phi} \approx 1$ Mpc (vertical dotted line) gives a sense of the physical distance.}
\label{fig:halos}
\end{figure*}

\begin{figure*}
\centering
\resizebox{88mm}{!}{\includegraphics{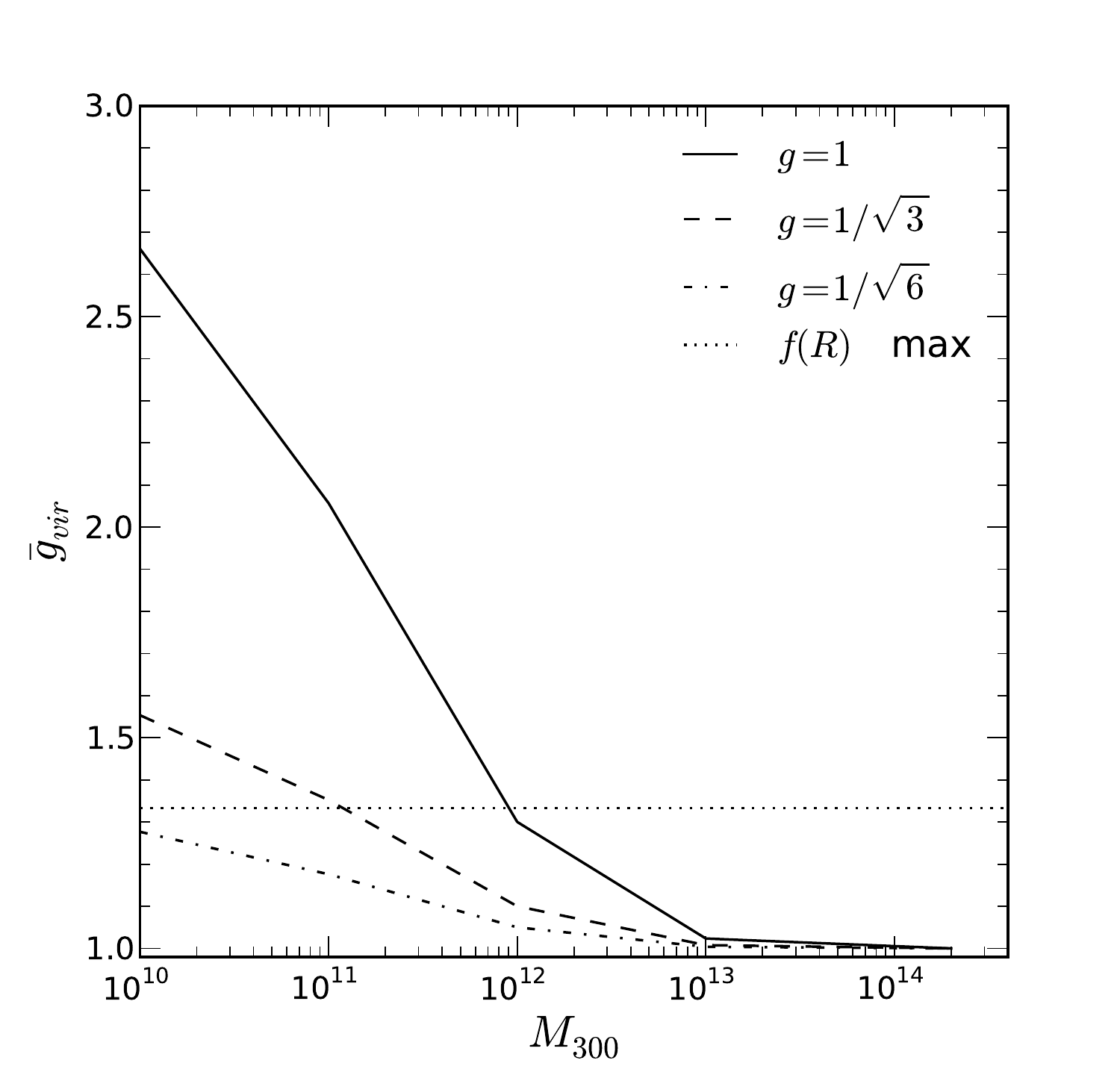}}
\resizebox{88mm}{!}{\includegraphics{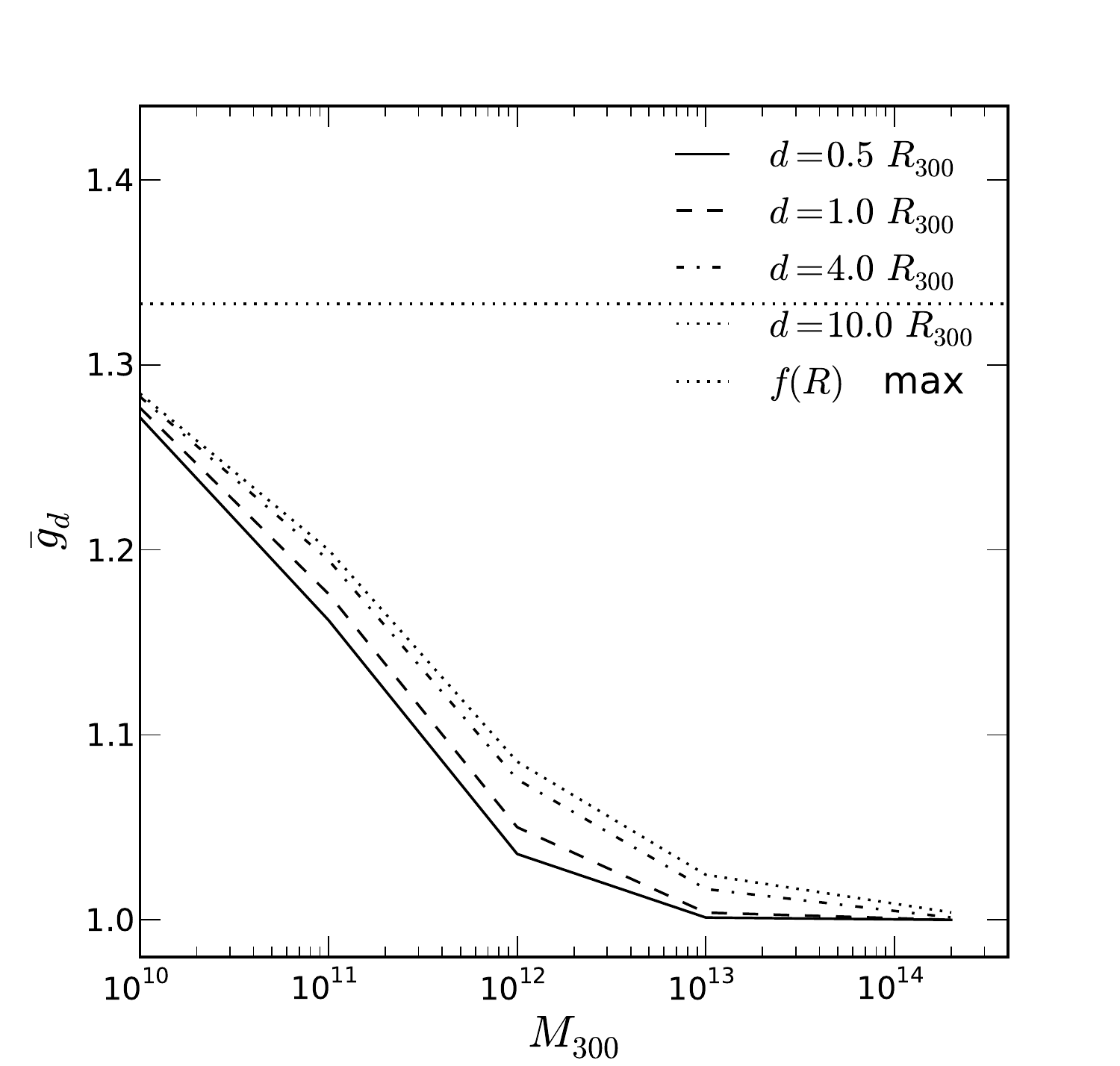}}
\caption{({\it left panel}): Averaged force deviation, $\gvir$, as a function of ``true'' or lensing mass for different values of the symmetry breaking field value, $\phi_0 = g M_{\rm s}^2 / M_{\rm Pl}$. ({\it right panel}): A modified averaged force deviation for the smallest field value $g = 1 / \sqrt{6}$. The average is taken over a distance $d$ from the center of the halo, with $d = 0.5, \; 1, \; 4,$ and $10 \; R_{300}$ from bottom. On both panels, the horizontal dotted line at  $4/3$ shows the maximum average for $f(R)$ modified gravity.}
\label{fig:gvir}
\end{figure*}

\begin{figure*}
\centering
\resizebox{78mm}{!}{\includegraphics[angle=0]{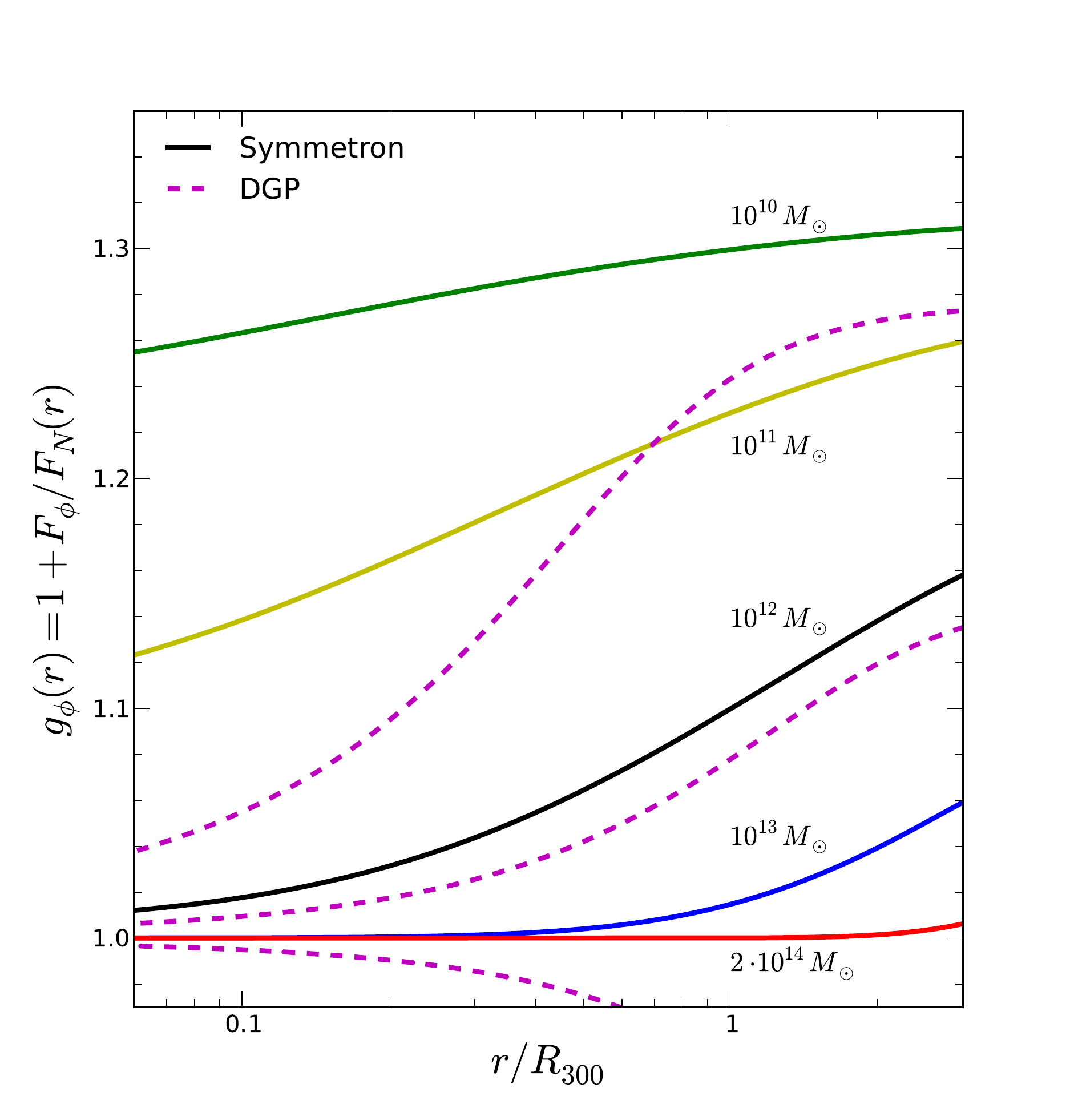}}
\resizebox{78mm}{!}{\includegraphics[angle=0]{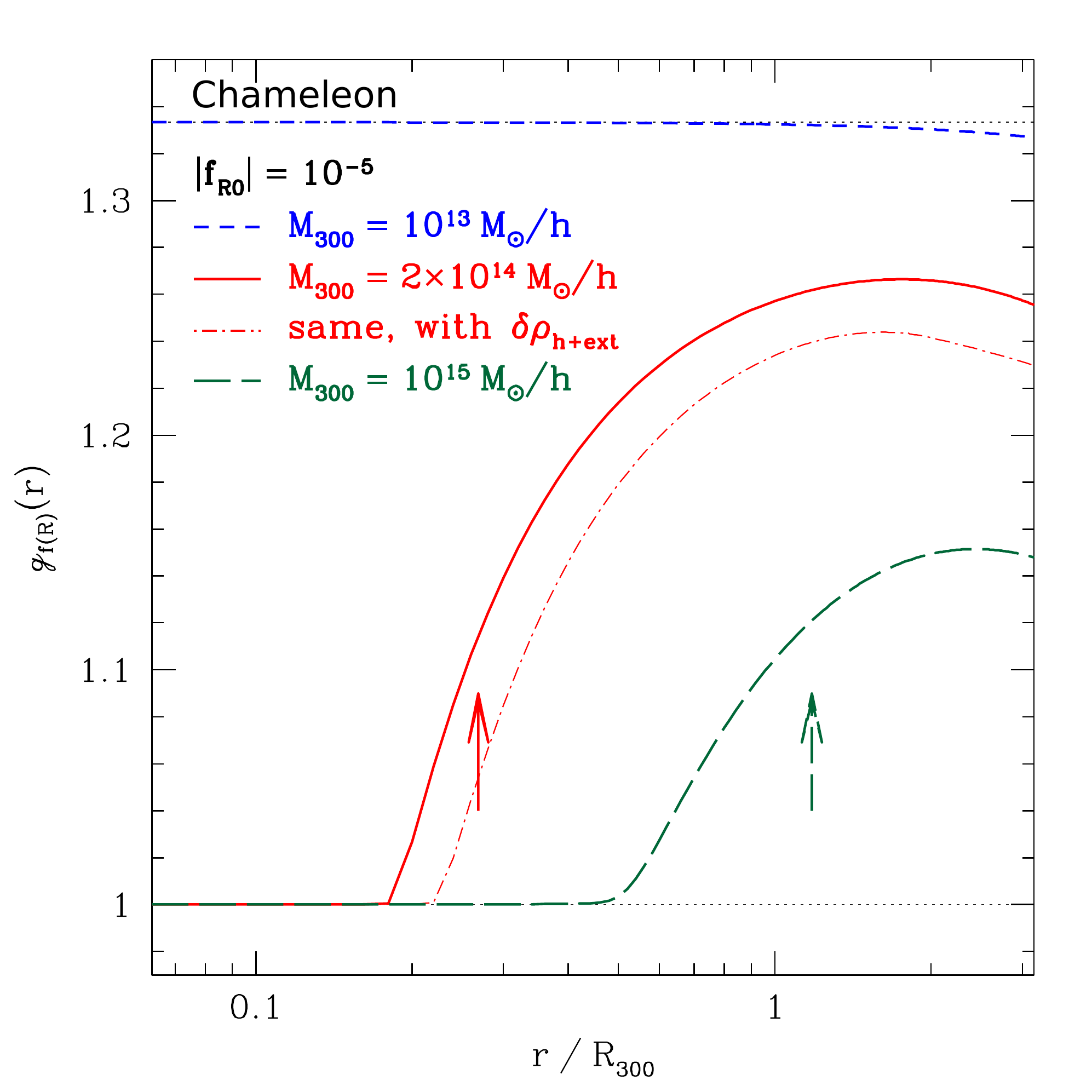}}
\caption{({\it left panel}): Here we plot $\g_{\phi} \equiv 1 + F_{\phi}/F_{\rm N}$ for comparison with $\g_{f(R)}$ in \cite{sch10}. Solid curves are from our symmetron model with $\phi_0 = M_{\rm s}^2 / \sqrt{6}M_{\rm Pl}$. Dashed curves are from normal branch DGP with crossover radius $r_c = 500$ Mpc, $r_c = 3100$ Mpc, and self-accelerating DGP, from top to bottom. ({\it right panel}): Reproduced from \cite{sch10}. Force deviation for Hu-Sawicki $f(R)$ around NFW halos. Arrows denote the radius where the chameleon thin-shell condition is first met. See \cite{sch10} for details of the DGP and $f(R)$ models. We see that the three types of screening predict distinct transitions in $F_{\phi}/F_{\rm N}$ with respect to both radius and mass (see text for details).}
\label{fig:compare}
\end{figure*}

Figure \ref{fig:halos} shows the force deviation $F_{\phi}/F_{\rm N}$ on a test particle given by Eq.~(\ref{eq:ratio2}) in the neighborhood of isolated NFW halos of various masses. Also pictured are the deviations for constant density spheres of the same mass. We plot using the symmetron vacuum value $\phi_0 = M_{\rm s}^2 / M_{\rm Pl}$, but as argued in Sec. \ref{sec:theory}, letting $\phi_0 \rightarrow g \phi_0$ simply shifts these curves down by a factor $g^2$. For all but the smallest halos ($10^{10} \msun$), the two profiles give significantly different results all the way out to the virial radius $R_{300}$.

In the left panel of Fig. \ref{fig:gvir}, we plot $\gvir$ given by Eq.~(\ref{eq:gvir}) for three values of the symmetron vacuum value: $\phi_0 = g M_{\rm s}^2 / M_{\rm Pl}$ with $g = 1, 1/\sqrt{3}, 1/\sqrt{6}$. Setting $g=1/\sqrt{6}$ fixes the max deviation at $4/3$ as in $f(R)$ and DGP theories. Equation~(\ref{eq:gvir}) implies $M_{{\rm dyn}, 300} = \gvir^{3/5} M_{300}$ so that the symmetron theory predicts, {\it e.g.}, the dynamical mass of a $10^{11} \msun$ halo to be 50\%, 25\%, or 10\% greater than the lensing mass for the three pictured values of $g$.

The right panel of Fig. \ref{fig:gvir} shows the same average taken out to larger radii $d \times R_{300}$, corresponding to 0.5, 1, 4, and 10 times the halo's virial radius. Although the peak deviations in Fig. \ref{fig:halos} are at approximately $7\; R_{300}$, the weighting by density in the integral of Eq.~(\ref{eq:gvir}) results in a relatively small increase in $\gvir$ with $d$ for a given halo.

In Figures \ref{fig:compare} we collect results for screening in the symmetron, DGP, and chameleon theories. The DGP and chameleon results are from \citep{sch10}. The left panel overlays our symmetron results (setting $g=1/ \sqrt{6}$) with three DGP models exhibiting Vainshtein screening. These are normal branch DGP with crossover radius $r_c = 500$ Mpc, $r_c = 3000$ Mpc, and self-accelerating DGP. The right panel of Fig. \ref{fig:compare} shows chameleon screening in the $f(R)$ model of \citep{huetal07} with field cosmological value $|f_{R0}| = 10^{-5}$. We refer the reader to \citep{sch10} for further details of these theories.

The ``thin-shell'' effect \citep{khoetal04} of the chameleon, in which only a thin shell at the edge of a screened object contributes to the fifth force, is evident in the rapid rise of the force deviations from zero to their peak at $r \approx 2 R_{300}$. In contrast, partially screened halos in the symmetron model show nonzero deviations at smaller radii that increase all the way to $\sim 7 R_{300}$ before declining back to zero (see Fig. \ref{fig:halos} for the large $r$ behavior). DGP models also show nonzero deviations that increase beyond $2 R_{300}$, eventually approaching a constant value.

The transition of the force deviations with halo mass also shows promise for distinguishing between the three screening mechanisms. For concreteness, consider the deviations at the virial radius $R_{300}$. For the chameleon background field value $|f_{R0}|=10^{-5} \; (10^{-6})$, a $10^{16} \msun \; (10^{15} \msun)$ halo shows no deviation while a $10^{13} \msun \; (10^{12} \msun)$ halo exhibits the maximum allowable deviation of 4/3. All smaller halos will likewise show a 4/3 deviation. In contrast, the symmetron $\fourteen \msun$ halo is completely screened at $R_{300}$ while the deviation is only maximized at 4/3 for dwarfs of mass $\lesssim 10^{9} \msun$. Thus, the degeneracy in the deviation exhibited between smaller halos in the chameleon model is not present in the symmetron. Furthermore, DGP models show no dependence on halo mass. These arguments imply that probes of modified forces spanning the mass range $10^{9}-10^{14} M_{\odot}$ would be effective at distinguishing between all three types of screening.

\section{Host-Satellite Effects}

\subsection{Model}

\begin{figure}
\includegraphics[width=0.7\columnwidth]{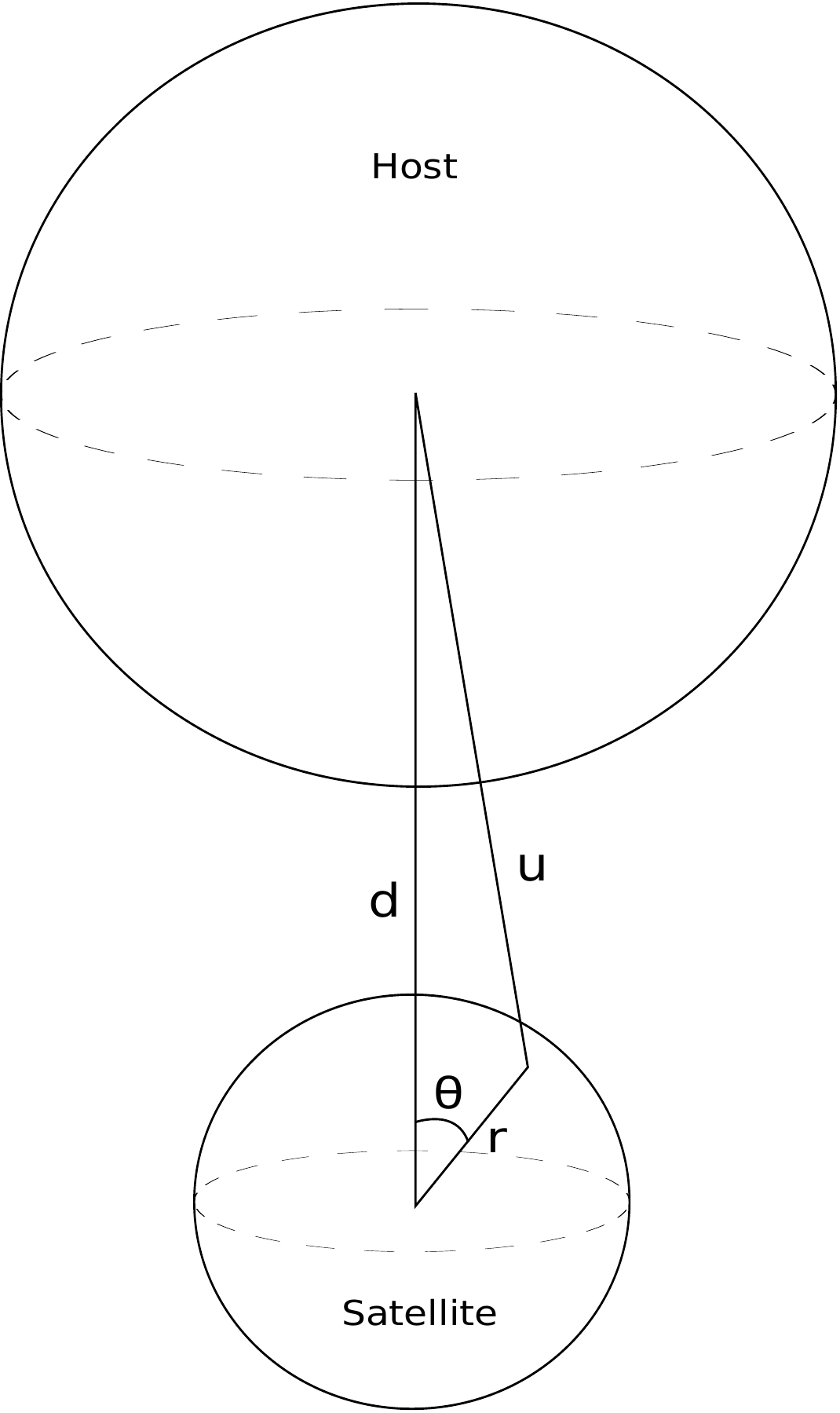}
\caption{We average the host density in spherical shells about the satellite's center. The parameter $d$ is the center-to-center distance between the halos, $r$ is the standard radial coordinate, $\theta$ is the azimuthal angle, and $u^2 = d^2 + r^2 - 2dr\cos{\theta}$.}
\label{fig:sketch}
\end{figure}

We now consider the force deviation that would be experienced by a test mass in the neighborhood of a satellite halo which is itself blanket screened by a nearby host. We model host and satellite with NFW profiles as before, determined by Eqs.~(\ref{eq:rho}), (\ref{eq:c}). To preserve spherical symmetry, we approximate the host profile by averaging its density in spherical shells around the satellite location. See Fig. \ref{fig:sketch}. Thus, for the average host profile we have
\bea \label{eq:host-density}
\langle\rho_{\rm h}\rangle(r) & = & \frac{1}{4\pi} \int {\rm d}\Omega \; \rho_{\rm host}(u(r,d,\theta)) \nonumber \\
 & = & \frac{1}{2} \int_0^{\pi} {\rm d}\theta \sin{\theta} \frac{4\rho_{\rm s}}{\frac{u}{r_{\rm s}}(1+\frac{u}{r_{\rm s}})^2} \nonumber \\
 & = & \begin{cases} \frac{4\rho_{\rm s}r_{\rm s}^3/d}{(d+r_{\rm s})^2 - r^2}, & 0 \le r \le d \\
  	\frac{4\rho_{0, {\rm s}}r_{\rm s}^3/r}{(r+r_{\rm s})^2 - d^2}, & d \le r \, , \end{cases}
\eea
where $u^2 = d^2 + r^2 - 2dr\cos{\theta}$, $d$ is the center-to-center distance between the halos, and $\theta$ is the azimuthal angle. Averaging the satellite density $\rho_{\rm sat}$ around its own center leaves its own NFW profile unchanged. We insert the total density
\be \label{eq:total-density}
\rho = \rho_{\rm sat} + \langle\rho_{\rm h}\rangle
\ee
in Eq.~(\ref{eq:new-eom}) with the boundary conditions~(\ref{eq:new-bc}), yielding the radial symmetron profile. The Newtonian force is given by integrating the profile Eq.~(\ref{eq:host-density}) in spherical shells to find the enclosed mass (see Appendix \ref{app:force}). We note that this averaged density profile will slightly overestimate (underestimate) the force deviation $F_{\phi}/F_{\rm N}$ on the side of the satellite nearer to (farther from) the host.

\begin{figure}
\includegraphics[width=0.88\columnwidth]{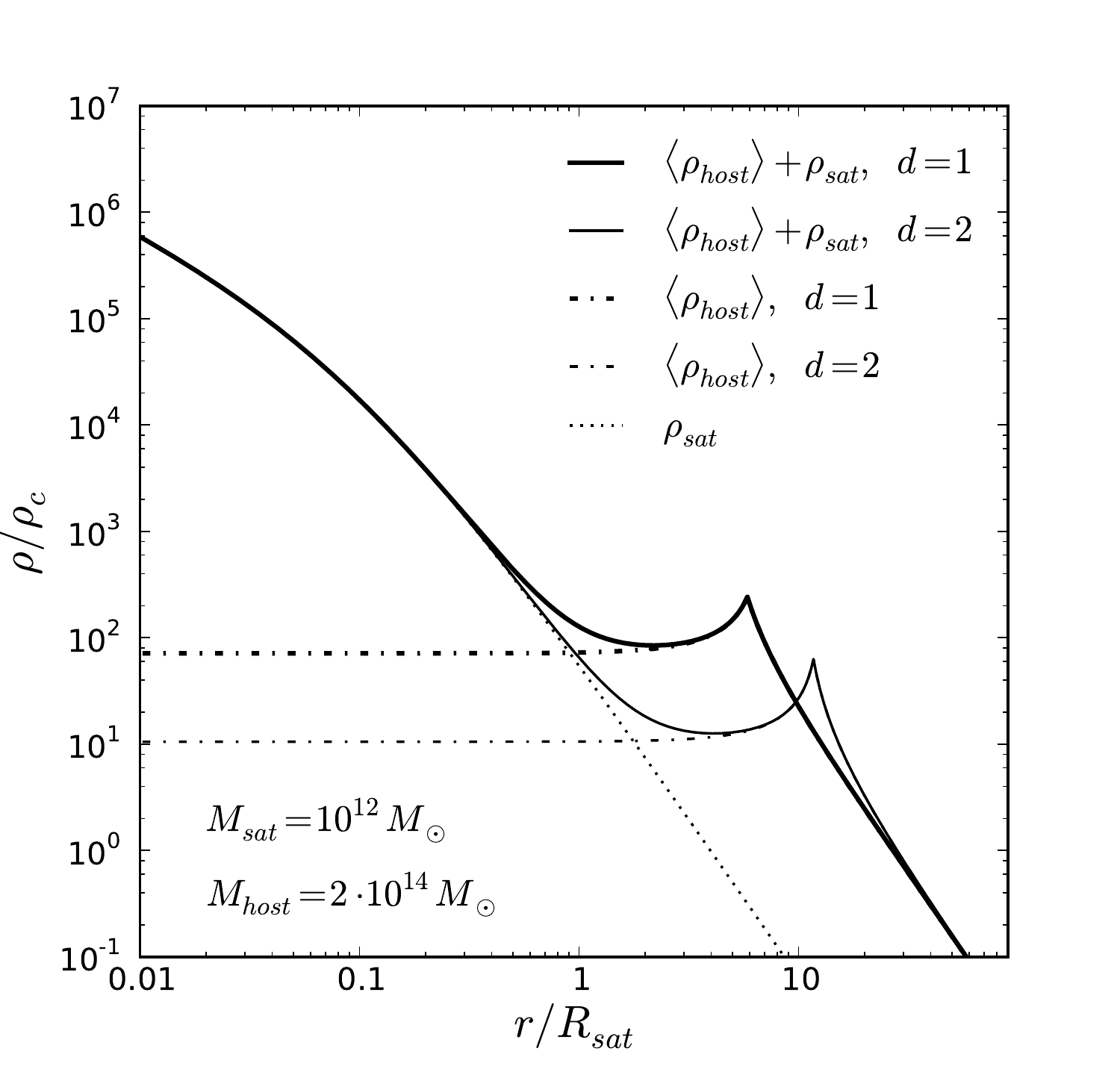}
\caption{Host-satellite density profile as a function of radius in units of the {\it satellite's} virial radius, centered at the satellite's location. Pictured are the NFW satellite profile (dotted), averaged host profile given by Eq.~(\ref{eq:host-density}) (dot-dashed), and the sum of host and satellite components (solid). The latter two curves are shown for host-satellite separation $d=1 R_{\rm host}$ (thick lines) and $d=2 R_{\rm host}$ (thin lines). In each case, the cusp in the total profile is located at the separation distance $d$. This plot assumes $M_{\rm sat} = 10^{12} M_{\odot}$ and $M_{\rm host} = 2\times 10^{14}$, although the profile is qualitatively similar for all relevant satellite and host masses. The host density becomes the dominant component at $r \sim R_{\rm sat}$.}
\label{fig:sat-d}
\end{figure}

Figure \ref{fig:sat-d} shows the density profile of a cluster-size ($2 \times 10^{14} M_{\odot}$) host modeled by Eq.~(\ref{eq:host-density}), an NFW satellite, and their sum, for two host-satellite separation $d= 1$ and $2 \; R_{\rm host}$. In the inner parts of the satellite its own NFW profile dominates the total density, while the host profile is a slowly increasing function of $r$. At $r \sim R_{\rm sat}$ the averaged host density becomes the dominant component. After the cusp, which occurs at the host center we see the host density rapidly transitions to the NFW $\rho \propto 1/r^3$ power law. The pictured density profile is qualitatively similar for all relevant satellite and host masses.

\subsection{Results}

\begin{figure*}
\centering
\includegraphics[width=6.8in]{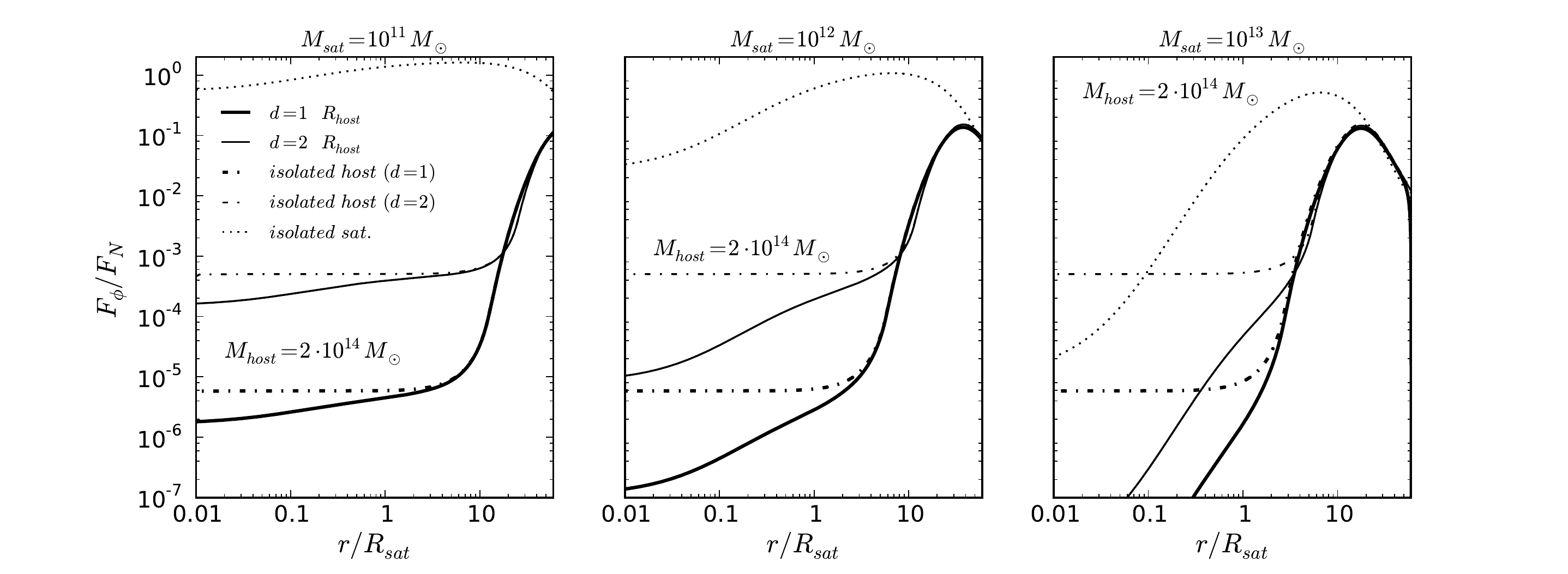}
\caption{Force deviation $F_{\phi}/F_{\rm N}$ for the total density profile of Eq.~(\ref{eq:total-density}) (solid lines). This is the deviation that would be experienced by a test mass in the neighborhood of a satellite halo which is itself blanket screened by a nearby host ($M_{\rm host} = \fourteen \msun$). The deviation is shown for two values of the host-satellite separation $d$: thick solid lines for $d=1 R_{\rm host}$ and thin solid lines for $d=2 R_{\rm host}$. Also shown are $F_{\phi}/F_{\rm N}$ for isolated halos with mass equal to the satellite (dotted) and host (thin and thick dot-dashed). The environmental screening from the host brings the modified forces of the satellite below 10\%, therefore nearly unobservable in each case. Figure \ref{fig:sat-f2} shows larger deviations for lower host masses.}
\label{fig:sat-f}
\end{figure*}

\begin{figure*}
\centering
\includegraphics[width=6.6in]{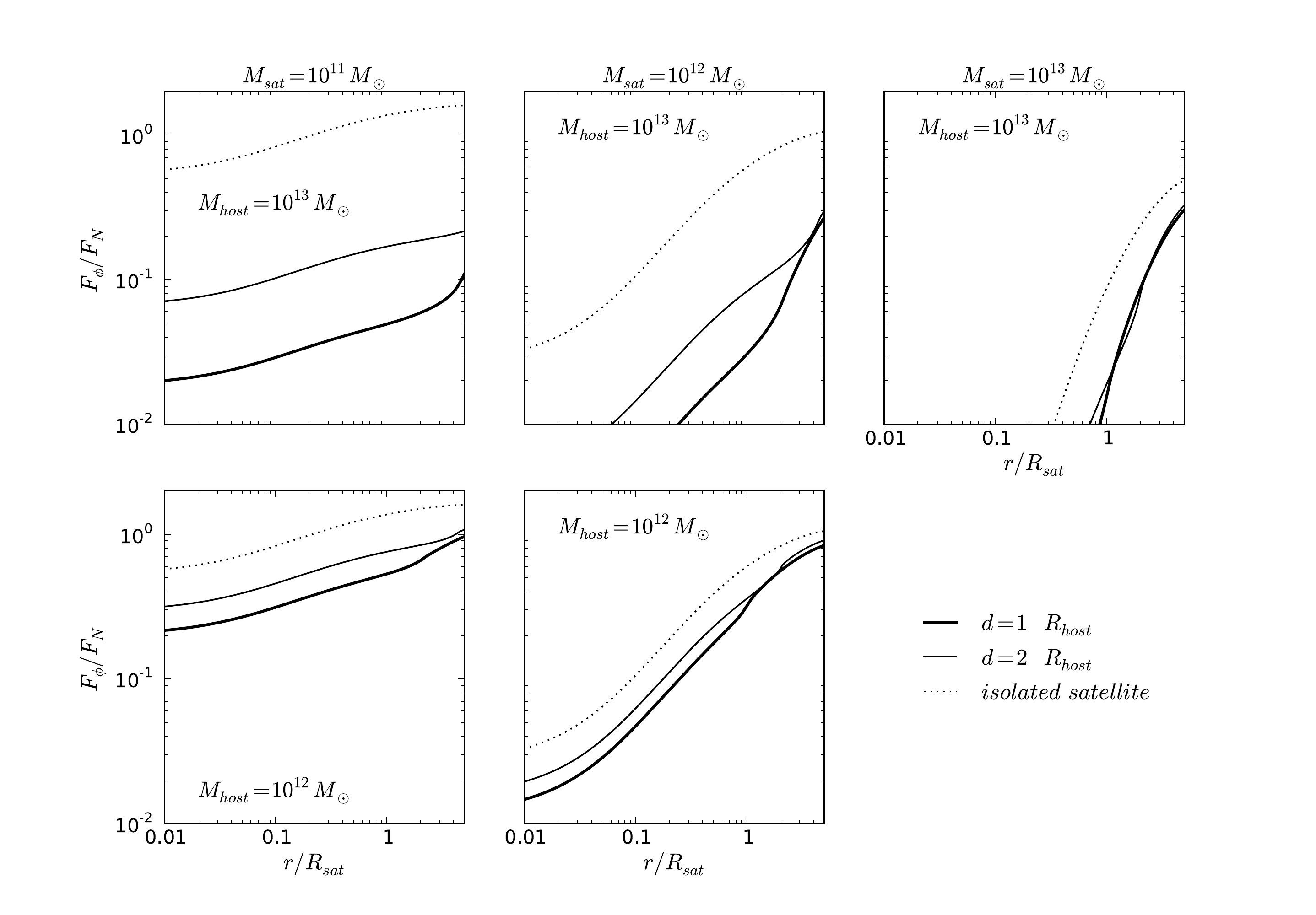}
\caption{Force deviation $F_{\phi}/F_{\rm N}$ for the total density profile of Eq.~(\ref{eq:total-density}) (solid lines), using smaller host masses than in Fig. \ref{fig:sat-f}. This is the deviation that would be experienced by a test mass in the neighborhood of a satellite halo which is itself (partially) blanket screened by a nearby host. The deviation is shown for two values of the host-satellite separation $d$: thick solid lines for $d=1 R_{\rm host}$ and thin solid lines for $d=2 R_{\rm host}$. For comparison, the dotted line plots the deviation for an isolated satellite: the difference between dotted and solid lines indicate the additional screening due to the nearby host. In each case, the host significantly decreases the force deviations of the satellite, but to a level that is still potentially observable.}
\label{fig:sat-f2}
\end{figure*}

The cluster's screening effect on the satellite is evident in Fig. \ref{fig:sat-f}. At low radii the force deviation $F_{\phi}/F_{\rm N}$ for the total density profile has a similar slope to that of the isolated satellite profile, but with an amplitude decreased below the observationally interesting level of $\sim 1$\%. In each panel there is a sharp transition of the solid line at the distance $d$ (i.e., location of the cusp in Fig. \ref{fig:sat-d}) separating the host and satellite. Here the force ratio for the total density profile is closely approximated by that of the host alone. At very large radii, all three curves fall off rapidly, as the symmetron force decays exponentially due to its finite compton wavelength. The environmental screening from the host brings the force deviation of the satellite below the level of observations for each case. Note that for this and the following two-halo calculations, we set $\phi_0 = M_{\rm s}^2 / M_{\rm Pl}$.

In order to see the partial screening of a satellite, it is necessary to consider smaller hosts. Fig. \ref{fig:sat-f2} shows the force deviation $F_{\phi}/F_{\rm N}$ for the same range of satellite masses and host satellite separations as in Fig. \ref{fig:sat-f}, but with host masses of $10^{13}$ and $10^{12} M_{\odot}$. For $M_{\rm sat} < M_{\rm host}$, including the host at $2 \; R_{\rm host}$ decreases the fifth force by 50-150\% relative to gravity, and halving the separation to $1 \; R_{\rm host}$ brings a further decrease of 5-10\%. For example, an isolated $10^{11} \msun$ has $F_{\phi} \approx 150\% F_{\rm N}$ at its virial radius, while if a $10^{12} \msun$ halo is 2 or 1 virial radii away, this deviation is cut down to $60\% F_{\rm N}$ or $45\% F_{\rm N}$, respectively. See Appendix~\ref{app:vary-d} for calculations of the modified forces at the satellite virial radius as a continuous function of $d$.

A few comments on the validity of the approximate host profile of Eq.~(\ref{eq:host-density}) and the results of Figs. \ref{fig:sat-f}, \ref{fig:sat-f2} are in order here. Since the NFW profile of the host does not change much across the diameter of the smaller satellite, we expect the force deviation predictions to be accurate in the regime $r \lesssim R_{\rm sat}$. Here our spherically symmetric approximation should slightly overestimate the deviation on the side of the satellite nearer the host, while underestimating the deviation on the far side. For $r \gg d$ the offset between host and satellite is negligible, so that Eq.~(\ref{eq:host-density}) approaches the NFW density profile Eq.~(\ref{eq:rho}). Thus, our approximation captures well the small and large $r$ behavior of the exact host-satellite system.

In the intermediate regime $r \sim d$ we note that the screening will vary widely between the near and far sides of the satellite. On the side nearer the host we should observe thorough blanket screening, while on the far side screening from the host may well be negligible, depending on the separation $d$. However, given that dynamical tracers of the satellite mass (such as stars, HI gas, and satellite galaxies) are generally confined to $r \lesssim R_{\rm sat}$, our model describes well the observationally relevant range.

\section{Discussion}

We have presented predictions for the modified forces of symmetron gravity around dark matter halos, modeling the halo density with the NFW profile. For a large range of halo masses $\ten - \thirteen \msun$ we find order unity deviations from GR at distances $1-7 \; R_{300}$ from the halo center, while the dwarf-size halos of $\ten - \eleven \msun$ exhibit large deviations throughout (Fig. \ref{fig:halos}).

We have also shown an average of this deviation given by the quantity $\gvir$ suggested by \citep{sch10}; our theoretical calculations of $\gvir$ (Fig. \ref{fig:gvir}) are simply related to observables by Eqs.~(\ref{eq:gvir}) and (\ref{eq:gvir-obs}). We find $\gtrsim 20\%$ differences over the mass range $\ten - \thirteen \msun$, indicating that observations of dynamical and lensing masses of galaxies are a promising way to constrain parameters of the theory.

Furthermore, $\gvir$ is necessary to test the mass function of the theory. Reference \citep{davetal11b} obtains the mass function from symmetron N-body simulations by counting halos as a function of their true or lensing mass. However, observations more commonly yield the dynamical mass of virialized structures; a means of converting between the two is therefore essential to constrain the theory using the mass function.

We have gathered predictions for symmetron, chameleon, and Vainshtein screening in the neighborhood of NFW haloes in order to find ways to distinguish between classes of MG models in realistic astrophysical situations. We find significant differences among the three screening mechanisms, including
\begin{itemize}
\item In contrast to the chameleon, the lack of a distinct thin-shell radius for the symmetron results in a nonzero deviation at small radii even for partially screened halos. Since visible tracers of galaxies are often well within the virial radius, the tests of \citep{huietal09, chaetal11, jaietal11} are more easily applied to test symmetron screening.
\item At a given radius, the chameleon deviations change rapidly with mass and therefore reach their maximum of 4/3 quickly. The symmetron exhibits more gradual changes in the deviation, while DGP models have no dependence on halo mass.
\end{itemize}

We note that there is an approximation involved in using the NFW profile: although this profile is a good fit to stacked halo densities in GR simulations, it is possible that the profile itself will change with the modified forces. However, the symmetron N-body simulations of \citep{davetal11b} are able to resolve halos of mass $\gtrsim 5 \times \twelve \msun$; for these larger halos we have compared our results for the symmetron field and find them qualitatively consistent with the N-body simulations. (Note that \citep{davetal11b} defines a parameter $z_{\rm SSB}$ related to $\mu$ and $M_{\rm s}$ (see \citep{davetal11b} for details) and focus on $z_{\rm SSB} = 2.0$. We have checked our results with theirs using this same value for $z_{\rm SSB}$, but our standard choice of parameters corresponds to $z_{\rm SSB} \approx 0.5$, for which the theory predicts relatively lesser deviations from GR. Thus, consistency with the simulations for the larger value of $z_{\rm SSB}$ is actually a {\it more} stringent test than consistency for our choice of parameters.)

This work has not considered two other screening mechanisms that have recently been discovered: the environmentally dependent dilaton \citep{dametal94, braetal10} and k-mouflage \citep{babetal09}. However, k-mouflage screening has been shown to be similar to Vainshtein (i.e., independent of halo mass), and due to the similar dependence of the derivative of the coupling ($\partial A / \partial \phi \propto \phi$) in symmetron and dilaton models, the dilaton screening may exhibit similarities to the symmetron.

We have also shown results for screening of satellite halos by larger neighboring hosts. We have approximated the host density by averaging its NFW profile in spherical shells about the satellite: the resulting profile is shown in Eq.~(\ref{eq:host-density}) and Fig. \ref{fig:sat-d}. We found that the environmental screening effect from a $\fourteen \msun$ cluster is sufficient to reduce deviations from GR well below the level of 1\% for smaller halos located within twice the host virial radius (Fig. \ref{fig:sat-f}). However, Fig. \ref{fig:sat-f2} shows that for smaller hosts of mass $\twelve - \thirteen \msun$ the force deviations around the satellite are decreased (relative to the isolated case) but still potentially observable. One caveat to this method should be mentioned: in reality, adding a second halo breaks the spherical symmetry of our system. Thus we expect that our results for two-body systems slightly underestimate (overestimate) the screening on the side of the halo nearer to (farther from) the second halo.

We comment briefly on the possibility of analytical solutions for two-body systems in symmetron modified gravity. There has been some success in finding approximate analytical solutions of two-body systems in chameleon theories. In \citep{motetal07} the thin shell effect is used to find solutions for uniform density spheres in a variety of configurations. Furthermore, \citep{pouetal11} expands these solutions to include an NFW host halo and test body (point mass) satellites of $M_{\rm sat} \le 10^{10} \msun$. However, these results do not straightforwardly translate to symmetron gravity: the thin-shell effect of the chameleon allows a clean division into 3 regions in the neighborhood of an object with a thin shell (inside the shell, the thin shell itself, and outside the shell); in contrast, the symmetron transitions more smoothly within screened objects (see Fig. \ref{fig:compare}), making such a partitioning much more difficult.

\section*{Acknowledgements}

We would like to thank Alex Borisov, Anna Cabr\'{e}, Yan-Chuan Cai, Michele Fontanini, Kurt Hinterbichler, Andrew Matas, Alan Meert, Jessie Taylor, and Vinu Vikram for many valuable discussions.

B.J. and J.C. are supported in part by NSF grant AST-0908027 and DOE grant DE-FG02-95ER40893. J.K. is supported in part by funds from the University of Pennsylvania, NSF grant PHY-0930521, and the Alfred P. Sloan Foundation.

\appendix
\section{Newtonian force of host}
\label{app:force}

Here we show the result of integrating the density profile of Eq.~(\ref{eq:host-density}) to find the Newtonian force of the host component. We have
\begin{widetext}
\bea
\frac{d\Psi_N}{dr} (r) & = & \frac{G}{r^2} M(<r) \nonumber \\
 & = & \frac{G}{r^2} 4\pi \int_0^r dr' r'^2 \langle\rho_{\rm h}\rangle(r') \nonumber \\
\frac{d\Psi_N}{dr} (r) & = & \frac{16\pi G\rho_0 r_{\rm s}^3}{r^2} \left[-\frac{r}{d} + \left(1 + \frac{r_{\rm s}}{d}\right) \ln{\left(\frac{r+d+r_{\rm s}-\sqrt{(d+r_{\rm s})^2-r^2}}{r-d-r_{\rm s}+\sqrt{(d+r_{\rm s})^2-r^2}}\right)} \right], \; \; r \le d \nonumber \\
 & = & \frac{16\pi G\rho_0 r_{\rm s}^3}{r^2} \left[-1 + \left(1 + \frac{r_{\rm s}}{d}\right) \ln{\left(\frac{2d+r_{\rm s}-\sqrt{2dr_{\rm s}+r_{\rm s}^2}}{-r_{\rm s}+\sqrt{2dr_{\rm s}+r_{\rm s}^2}}\right)} \right. \nonumber \\
 & & \left. + \frac{1}{2}\left(1-\frac{r_{\rm s}}{d}\right) \ln{\left(\frac{r+r_{\rm s}-d}{r_{\rm s}}\right)} + \frac{1}{2}\left(1+\frac{r_{\rm s}}{d}\right) \ln{\left(\frac{r+r_{\rm s}+d}{2d+r_{\rm s}}\right)}  \right], \; \; d \le r \, .
\eea
\end{widetext}

\section{Variations of host-satellite separation}
\label{app:vary-d}

Although using the averaged host profile of Eq.~(\ref{eq:host-density}) preserves spherical symmetry, we must still solve the nonlinear Eq.~(\ref{eq:new-eom}) for each choice of host-satellite separation $d$. To see the effect of varying this parameter continuously, we make a different approximation. If the symmetron profile of the host at the location of the satellite has value $\phi_{\rm host}(d)$, and this value varies little across the diameter of the satellite, we can instead solve the approximately equivalent system of an isolated satellite with asymptotic field value $\phi (r \rightarrow \infty) = \phi_{\rm host}(d)$.

\begin{figure*}
\centering
\resizebox{85mm}{!}{\includegraphics{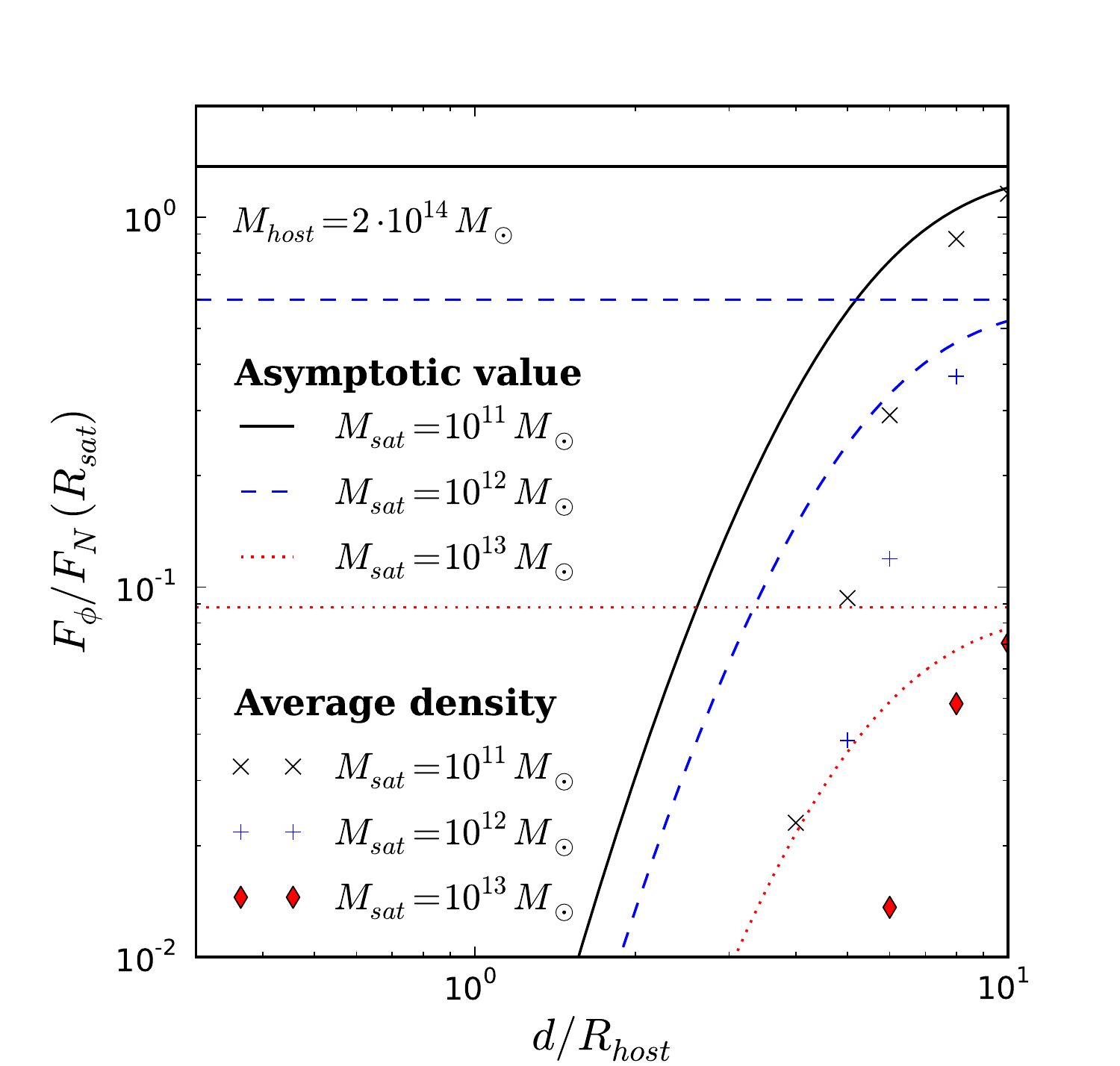}}
\resizebox{85mm}{!}{\includegraphics{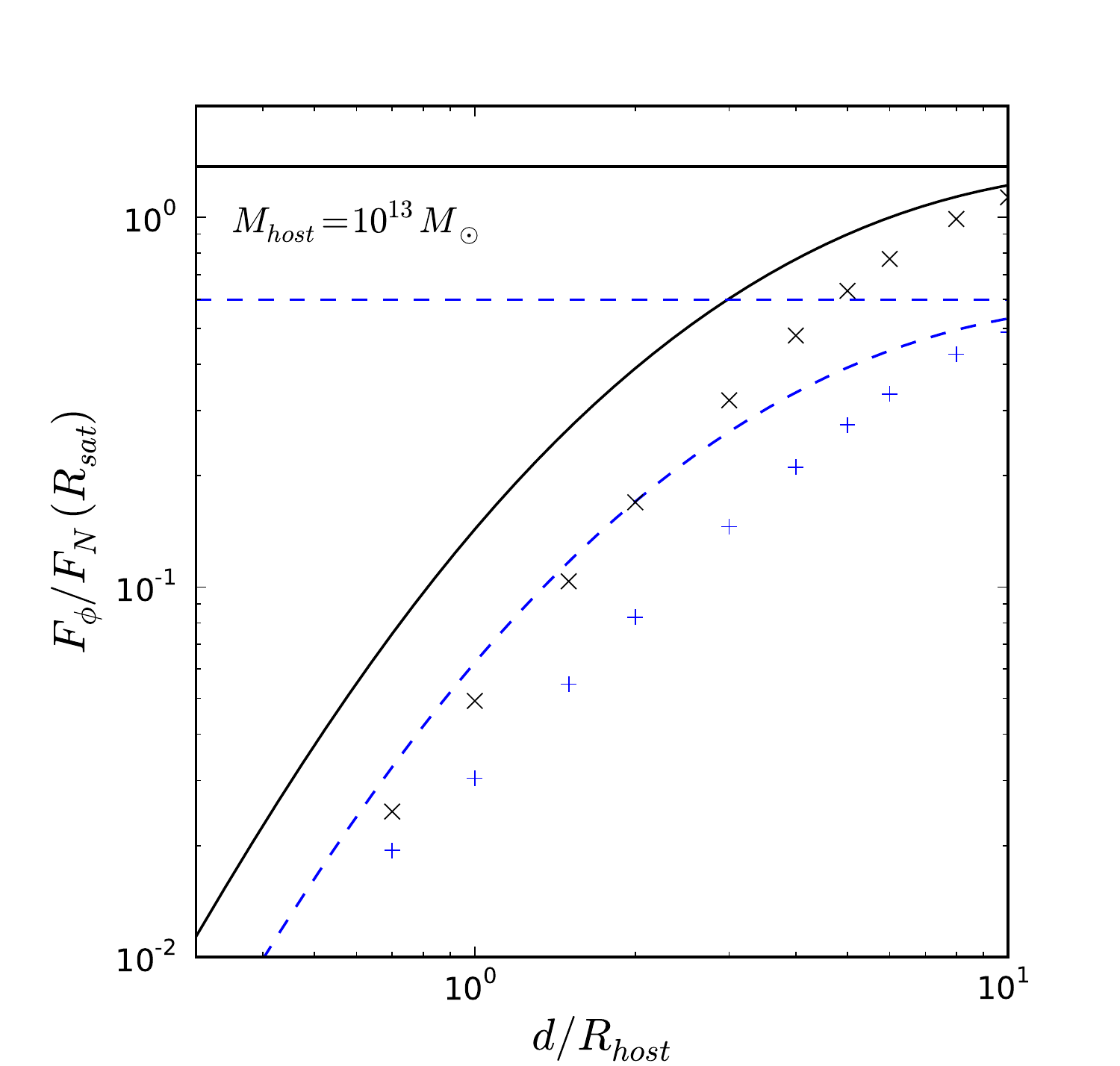}}
\caption{({\it left panel}): Satellite force deviation $F_{\phi}/F_{\rm N}$ at $1$ $R_{\rm sat}$ as a function of center-to-center distance $d$ from a nearby host halo. Continuous lines display the result using the asymptotic value of the field appropriate for a relatively small object at distance $d$ from the host (see text for details). Satellite masses $M_{\rm sat} = \eleven, \twelve, \thirteen \msun$ correspond to solid, dashed, and dotted curves, respectively. Also shown for the same range of masses are the results using the total density profile of Eq.~(\ref{eq:host-density}) applied at a discrete set of points. Horizontal lines show the force deviation of isolated satellites. The host mass is $\fourteen \msun$. For the $\eleven$ and $\twelve \msun$ satellites, the screening from the host has a significant ($\sim 10\%$) effect even at separations of $\approx 7-8 \; R_{\rm host}$. ({\it right panel}): Same as left, but for a host mass of $\thirteen \msun$. }
\label{fig:separation}
\end{figure*}

Fig. \ref{fig:separation} plots the satellite force deviation at $1 R_{\rm sat}$ as a function of $d$. This is done using two different approximations: the method using the asymptotic value of the field appropriate for a small object at distance $d$ from the host and the total density profile method of Eq.~(\ref{eq:host-density}). Plotting continuous curves for the latter is impractical: the shooting method must be employed to solve Eq.~(\ref{eq:new-eom}) for each distance $d$. Fortunately, a discrete set of points shows the trend sufficiently well.

We see that the computationally much simpler asymptotic value method approximates well the more sophisticated average density calculation as long as the host and satellite masses are within 2 orders of magnitude. Here the simpler method overestimates the fifth force by only $5-30\% \; F_{\rm N}$ over the range of separations considered. The differences exceed this level for the case of a $10^{11} \msun$ halo in the neighborhood of a $2 \times 10^{14} \msun$ host, but only at separations of 5-6 $R_{\rm host}$ where it reaches $\approx 50\%$.

\bibliography{sources}

\end{document}